# Simulation Based Probabilistic Risk Assessment (SIMPRA): Risk Based Design

Hamed S Nejad

*Center for Risk and Reliability, 0151 Glenn L. Martin Hall, University of Maryland, College Park, MD 20742, USA*

Tarannom Parhizkar

*B. John Garrick Institute for the Risk Sciences, University of California, Los Angeles, 404 Westwood Plaza, Los Angeles, CA 90095, USA. E-mail: tparhizkar@ucla.edu*

Ali Mosleh

*B. John Garrick Institute for the Risk Sciences, University of California, Los Angeles, 404 Westwood Plaza, Los Angeles, CA 90095, USA. E-mail: Mosleh@ucla.edu*

The classical approach to design a system is based on a deterministic perspective where the assumption is that the system and its environment are fully predictable, and their behaviour is completely known to the designer. Although this approach may work fairly well for regular design problems, it is not satisfactory for the design of highly sensitive and complex systems where significant resources and even lives are at risk. In addition it can results in extra costs of over-designing for the sake of safety and reliability.

In this paper, a risk-based design framework using Simulation Based Probabilistic Risk Assessment (SIMPRA) methodology is proposed. SIMPRA allows the designer to use the knowledge that can be expected to exist at the design stage to identify how deviations can occur; and then apply these high-level scenarios to a rich simulation model of the system to generate detailed scenarios and identify the probability and consequences of these scenarios. SIMPRA has three main modules including Simulator, Planner and Scheduler, and it approach is much more efficient in covering the large space of possible scenarios as compared with, for example, biased Monte Carlo simulations because of the Planner module which uses engineering knowledge to guide the simulation process.

The value-added of this approach is that it enables the designer to observe system behaviour under many different conditions. This process will lead to a risk-informed design in which the risk of negative consequences is either eliminated entirely or reduced to an acceptable range. For illustrative purposes, an earth observation satellite system example is introduced.

*Keywords*: Simulation based probabilistic risk assessment, complex systems, risk-based design, dynamic risk assessment, planner, earth observation satellite, scenario planning, risk-informed design, reliability.

## 1. Introduction

The classical approach to design is based on a deterministic perspective where the assumption is that the system and its environment are fully predictable, and their behaviour is completely known to the designer, (Parhizkar, Aramoun and Saboohi, Efficient health monitoring of buildings using failure modes and effects analysis case study: Air handling unit system. 2020) (Ezzatneshan 2014) (Parhizkar, Aramoun and Esbati, et al. 2019). Input variables are assumed to be fixed and safety factors are applied to ensure the reliability and robustness of the design (Mahadevan, Smith and Zang, System risk assessment and allocation in conceptual design. 2003) (Parhizkar, Balali, and Mosleh, An entropy based bayesian network framework for system health monitoring. 2018). Although this approach may work fairly well for regular design problems—given, of course, the possible extra costs of over-designing for the sake of safety and reliability—it is not satisfactory for the design of highly sensitive and complex systems where significant resources and even lives are at risk.

In last decades, risk based design has been the focus of attention for different applications such as ships (Kujala, et al. 2019) (Tan, Tao and Konovessis 2019) (Garbatov, Sisci and Ventura 2018), wearable sensors (Arpaia, et al. 2021), infrastructure design (Byun and Hamlet 2020), pipelines (Hasan, et al. 2018), and nuclear power plants (Andreev, et al. 1998) (Schumock, et al. 2020). The risk assessment models utilized in risk based design problems can be categorized as qualitative (Gazourian, et al. 2021) and quantitative methods. Qualitative methods are mostly used in applications with limited data/knowledge about the system behaviour or when the goal of the modelling is to provide a comparison between possible design scenarios. However, quantitative methods provide the risk level of a design that could be compared with other designs or used to verify system meets design safety requirements. Quantitative risk assessment (QRA) or probabilistic risk assessment (PRA) is a systematic risk analysis approach to quantifying the risks associated with the design and operation of an engineering system (Parhizkar, Hogenboom and Vinnem, Data driven approach to risk management and decision support for dynamic positioning systems. 2020) (T. Parhizkar, J. E. Vinnem, et al. 2020) (T. Parhizkar, I. B. Utne, et al. 2021) (T. Parhizkar, I. B. Utne, et al. 2021). The dynamic characteristics of an engineering system, stochastic processes, sequential dependencies of equipment or components, inspection and testing time intervals, aging of equipment or components, and seasonal changes, affect system PRAs significantly (Parhizkar, Mosleh and Roshandel, Aging based optimal scheduling framework for power plants using equivalent operating hour approach. 2017) (T. Parhizkar, J. Vinnem, et al. 2020) (Sotoodeh, et al. 2019).

The conventional PRA methodologies includes, but not limited to fault trees, Bayesian network, Marcov models, Fuzzy theory, Monte Carlo methods, and event sequence diagrams. These methods have limited capacity for quantifying the dynamic behaviour of complex systems. In recent years, a new approach of dynamic probabilistic risk assessment (DPRA) is introduced that can address time-dependent effects in PRAs and provide precise risk estimations resulting from system interdependencies and dynamic behavior.

Dynamic PRA (DPRA) methods consider time-dependent interdependencies among system components, including technical, environmental, and organizational factors, (Hogenboom, Parhizkar and Vinnem 2021). The common DPRA methods include dynamic event tree (Zhou , Reniers and Khakzad 2016) (Swaminathan and Smidts 1999), dynamic fault tree (Vesely, et al. 1981) (Gascard and Simeu-Abazi 2018), dynamic Bayesian network (BN) methods (Y. Lei and Lee 2012) (Sterritt, et al. n.d.), and hybrid methods (Mosleh 2014) (T. Parhizkar, I. Utne, et al. 2021). Generally, these methods generate operational scenarios of complex systems through time. the main challenge of these methods is that the number of possible scenarios after an incident increases significantly. This challenge is explained in more detail through a case study in (T. Parhizkar, J. Vinnem, et al. 2020) (T. Parhizkar, I. Utne, et al. 2021).

In this paper, a new simulation based method is proposed that guides the scenario generation in DPRAs. The simulation based probabilistic risk assessment (SIMPRA) methodology, can be seen as a combination of continuous event tree (CET) and discrete dynamic event tree approaches (DDET). In fact, SIMPRA is a CET approach that is guided by a planner that, like DDET approaches, takes advantage of the classical event sequence diagram and event tree approaches with binary logic restrictions removed (Hu, Parhizkar and Mosleh 2021). In SIMPRA, events are simulated in continuous time. When the simulation reaches a branch point, the direction that it has to take is decided based not only on a random selection of events according to the probability of branches, but also by the importance measure of events and the expected entropy gain from simulating each branch. The SIMPRA methodology is explained in more detail in (Hu, Parhizkar and Mosleh 2021).

In Section 2, a review on the SIMPRA methodology is presented. Section 3 presents risk acceptance and risk management methodology based on the SIMPRA method.

In order to explain the method, An earth observation satellite system is presented in Section 4 as an example. Lastly, concluding remarks are presented in Section 5.

## 2. SIMPRA Method

In practice, risk assessment is performed by first identifying how a system might deviate from its intended performance, second deciding how probable these deviations are and third determining what the consequences of these deviations might be (Kaplan, Haimes and Garrick 2001). SIMPRA allows the designer to use the knowledge that can be expected to exist at the design stage—which is based on the mapping of functional requirements in the functional domain to the design parameters in the physical domain (Suh 2001)—to:
a) generate high level risk scenarios, that is, to identify how deviations can occur; and then
b) apply these high-level scenarios to a rich simulation model of the system to generate detailed scenarios and identify the probability and consequences of these scenarios.

As discussed in the introduction section, DPRA is usually interpreted as an exploration of the space of possible event sequences to gain risk information. A number of simulation methods have been used in DPRA to help the analyst understand the behaviour of the system under a variety of conditions (Mahadevan and Raghothamachar, Adaptive simulation for system reliability analysis of large structures. 2000) (Marseguerra, et al. 2003) especially those leading to risky outcomes. However, the SIMPRA approach is much more efficient in covering the large space of possible scenarios as compared with, for example, biased Monte Carlo simulations because of the planner element which uses engineering knowledge to guide the simulation process. This allows SIMPRA to avoid the slow convergence of most biased Monte Carlo methods which aim at finding an optimal sampling function while disregarding the structure of the system under investigation.

Figure 1 presents the main components of the SIMPRA methodology including simulator, scheduler and planner (Hu, Parhizkar and Mosleh 2021). In the following main components are explained briefly. A detailed description of SIMPRA method is presented in (Hu, Parhizkar and Mosleh 2021).

**Simulator:** is responsible for generating detailed system behaviour as intended by system designers. Simulation consists of models to simulate the behaviour of software, hardware or human components of the problem under study.

**Schedular:** manages the simulation process by saving system states, deciding the branch selection, adjusting the simulation levels of detail, and restarting the simulation. The scheduler guides the simulation toward the generated plan by the planner. The scheduler keeps track of the simulation and guides it adaptively.

**Planner:** uses engineering knowledge to generate high level scenarios. The planner guides the simulation by providing high-level scenarios to the scheduler. The generated scenarios could be incomplete or incorrect. The planner learns the detailed simulation results after a number of simulation runs (Figure 1: link IV) and provides suggestions for the system analysts on possible ways to enhance the plan model (Figure 1: link II) for instance by generating unseen scenarios or removing impossible ones.

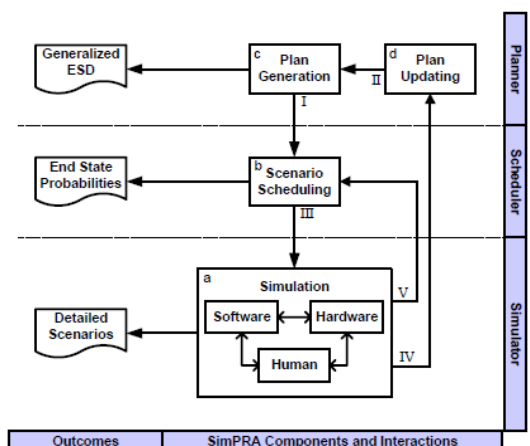

Figure 1. SIMPRA framework components and the interactions among them (Hu, Parhizkar and Mosleh 2021)

*2.1. Simulator*

SIMPRA relies on an executable model of the system, which emulates system behaviour. Given the operational profile and adequate input, we assume that this model would reproduce the behaviour of the real system under specific circumstances. The model is an abstraction of the real system. The abstract level may change during the course of the risk analysis. When we have a better understanding of the system, or at a later stage of system development, we may replace the existing model with a more detailed model.

In a typical DPRA problem, the system under investigation consists of discrete component state and continuous process variables. The evolution of continuous process variables is governed by deterministic physical and logical laws. Physical processes are described using mathematical expressions of such laws. The components have discrete states, which represent a set of operational modes or configurations. The component state may change from one state to another. This change may be internally determined by system logic and/or physical laws, or described by statistical or probabilistic laws. The occurrence of discrete state transition is defined as an event. The change of component state may in turn influence the mathematical equations describing their behaviour in each state. Examples of stochastic events include component malfunction with time-to-failure following a Weibull distribution. One example of deterministic event is a fuse melting in a circuit when the electrical current reaches a certain level. The model is a combination of both deterministic and stochastic models. The stochastic process dynamically interacts with the continuous-time, deterministic processes. The interactions between them consist of at least:

- The state of the discrete system determines the boundary conditions of the continuous-time process;
- The continuous-time process, e.g. system dynamics, such as pressure or temperature, may affect the stochastic process, e.g. failure rate;
- The continuous-time process may generate events, which in turn change the discrete state of the system.

The simulation model consists of hardware, software and human crew. Modelling choices for hardware systems are well established in most cases. The evolution of the model is traced by solving the continuous-time system in intervals between discrete events generated by the discrete system. Whenever the time for a scheduled event is reached, the continuous-time simulation is stopped, and the corresponding event is executed. In some cases, the continuous-time system may generate an event, e.g. one of the variables crosses a given threshold.

*2.2. Scheduler*

There are two types of systematic exploration supported in SIMPRA. The first type is a full-scale systematic exploration. The second type is a systematic search for specific events. SIMPRA also supports the systematic exploration strategy as in discrete dynamic event tree (DDET) methods. Once a branch point is reached and branches are proposed to the scheduler, the scheduler would explore all proposed branches. The exploration of event sequences is managed in a depth-first manner, as in ADS (Chang and Mosleh 2007).

At the branch point, the current system state at the branch point and all the branches are stored in a database, and the first branch is executed, and the simulation continues. When an end state is reached, the scheduler goes one step back to the previous branch point. If at least one branch still remains to be explored, the scheduler would retrieve the state of the branch point to re-initialize the simulation and explore the new branch. The simulation is restarted until another end state is reached. When all possible branches of that branch point have been explored, the scheduler brings the simulator one step backwards, until the dynamic tree exploration is completed.

In addition, there is a user-defined parameter limitation (Plim). If the probability of branch gets lower than Plim, that branch is no longer simulated. The probabilities of such event sequences are collected as truncated in the scheduler. We want to keep the sum of the truncated probabilities low; otherwise, it may introduce significant error in the estimator.

If we are performing a full-scale systematic search, the event sequences are generated and explored in a systematic way. Users define the number of event sequences they want to generate. The number of event sequences generated by the exploration of one round is limited by the number of branch points generated during the simulation, and maybe lower than the user defined number. The systematic search will go on for another round until the number of event sequences is greater than that designated by the user.

*2.3. Planner*

Plan is a map to guide the exploration. The planner is a module of SIMPRA to generate such a map. Planner collects useful knowledge about the contributors to different classes of risk scenarios and generates simulation roadmaps. In using the planner, the first step is for the users to construct an abstract model of the system. The abstract model consists of a component tree, a functionality tree, and a state transition diagram. The component tree is used to establish relationships among various sub-systems and components involved in the system. A sub-component node can further have sub-components. The relation among components and subcomponents are presented by AND and OR gates. The 'AND' relation shows the distinct elements while 'OR' is used to present the redundancies.

The functionality tree is used to establish the relationships between the various sub-processes and functionalities associated with the system. Based on the component and functionality trees, component-functionality matrix could be developed. In this matrix, functionalities are defined by the state of the components that are needed to perform the tasks. It is assumed that components have binary states; they either work or fail. Since we are interested in the functionalities that change the state of the system, component failures are only seen in this context. The behaviour of the system is defined by changes in the state of the system and its elements in the hierarchy. Changes in the states are triggered (or from the planner's point of view are initiated) by changes in the functionalities and scenarios are generated by putting these changes in order to reach the goal states.

The state transition diagram uses the explicit knowledge about the system to determine the possible state transitions in the system. Explicit knowledge about the system behaviour could be obtained by Finite-State-Machines (FSM). The state transition diagram is used to draw the required state-relationship, associate them with the selected components and functionalities from the component-functionality matrix, and then generate a plan (Nejad-Hosseinian 2007). The component tree and functionality tree are stored in a database file. The planner would query the database file while generating the plans.

### 2.4. Scenario exploration

The knowledge of the system vulnerabilities can be expressed as a list of scenarios which may lead to undesirable end states. Recalling the definition of scenario, the scenario does not have to be complete event sequences. There is no requirement that the list of scenarios would cover all the event sequence space. In event tree analysis, the complete event sequences need to be laid out by the analyst, and it is essential to accurately cover all the event sequence space with event trees. Fulfilment of this requirement of the event tree relies on the expertise of the risk analysts, which is hard to verify, especially when dealing with new systems where we do not have much relevant experience. By contrast, in the proposed DPRA methodology, scenarios listed in a plan serve as a guide for the simulation. Due to the randomness of the simulation, it is believed that with a large number of guided simulation stories, all event sequences of interest would be touched.

The random elements of the simulation would be controlled by the scheduler. The objective of the scheduler is to distribute the simulation effort among different scenarios. The scenario with high importance would be explored with higher priority, while all other scenarios also have a chance to be simulated. Among the important scenarios, we want the scheduler to guide the simulation efforts evenly. Simulation focused on only one or few scenarios, leaving other important scenarios untouched, is undesirable. Among the desired properties, the scenario exploration in this methodology will:

- Maintain sufficient coverage of important scenarios
- Guide simulation toward areas of greatest uncertainty
- Continuously adjust priorities based on simulated results
- Avoid test areas known to definitely lead to a specific end state
- Guide simulation to be able to cover all the event sequence space

### 3. Risk Based Design

The SIMPRA framework can be used for the risk assessment of a design to identify the worst case scenarios and the probability of their occurrence as well as the total risk of the system's behaviour under uncertain conditions. The value-added of this approach is that it enables the designer to observe system behaviour under many different conditions. The designer can also modify the design to compare the results of the risk assessment under different design specifications. This process will lead to a risk-informed design in which the risk of negative consequences are either eliminated entirely or reduced to an acceptable range. The overall structure of the SIMPRA approach to risk-based design is illustrated in Figure 2. The first step—System Modelling—is to create a virtual model of the system with the appropriate level of detail. The second step—Knowledge Acquisition for Scenario Generation—is to acquire additional information about the system and its environment that will be used to automatically generate risk scenarios in the third step—Risk Scenario Generation. The fourth step—Simulation—consists of the actual running of the simulation, the results of which are reported as the probabilities of the end-states as well as the worst case scenarios in the fifth step—Risk Assessment. In the sixth step—Risk Acceptable—the designer or risk analyst makes a determination as to whether or not the risks associated with the scenarios and end-states are acceptable. If the risks are not acceptable and must be reduced through changes in the design of the system, those changes are applied to the simulation model (and to the planner model if necessary) in the seventh step—Design Risk Management. The risk assessment process is repeated as many times as necessary and ends only when the designer is satisfied with the level of risk in the system or has found an acceptable range of behaviour for each component of the system that does not jeopardize the robustness of the overall system's behaviour.

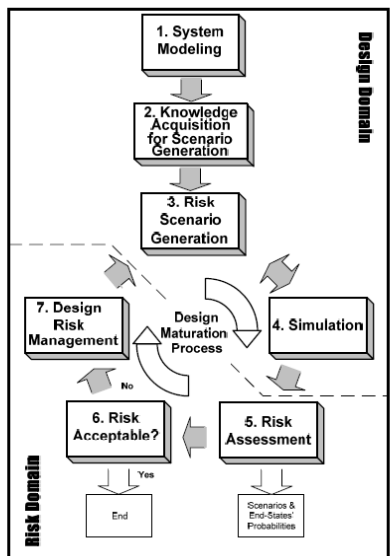

Figure 2. Overall structure of SIMPRA approach to risk-based design

## 4. Risk Acceptance and Risk Management

The level of acceptable risk is usually defined by the requirements. Any deviation from the expected functionality and behavior of a system can lead to an unwanted consequence and needs to be considered in the risk assessment process. Safety, robustness, dependability and reliability are all some of the characteristics that can be affected by changes in the behaviour of a system and can be evaluated through the simulation process.

When studying the life-cycle of a component in a system that has triggered an accident, it is usually possible to distinguish four time intervals:

1. Assuming that the component was not defective when it was initially put into the system, there is a time frame in which the component does its job properly and within the range of expectations. This time frame can be named the "productive time".
2. The second time frame is that in which the component has not yet failed to perform its function but is pushing toward the edges of acceptable performance range. This is usually accompanied by some sorts of changes in the system behaviour that can signal an observer that something has changed in the system. This time frame is called the "degradation time"
3. The third time frame is when the component finally fails to perform its job and the system starts moving towards a failure. This is called the "failure time".
4. Finally, most failures can become triggering events for additional series of failures if their damage is not properly controlled. The time it takes to control the effects of a failure on its environment is called the "failure recovery and management time".

A familiar example of the failure life-cycle can be seen in the performance of automobile brakes. When a car's brake pads are new, they usually work very well for a long time. When they begin to wear out, they make a squeaking noise when the driver tries to stop the car. This is the degradation time during which the brakes are still working but are pushing toward the edges of acceptable performance range. If not soon replaced, the brakes will soon stop working altogether and may cause a serious accident. In the event that a car accident does occur, the recovery and management time may be devoted to ensuring the safety of other cars in the area.

Different risk management tools are designed to work within different time frames of the failure life cycle and help to prevent or at least minimize the consequences of the failure or the recovery effort that follows. These risk management tools can be divided into 7 categories:

1. Life design: This is the most common approach to risk management. If the "productive time" of a component is short, the component can be replaced with a more reliable one. Alternatively, the conditions under which the component is performing may be improved so that it experiences less stress while performing its job.

2. Maintenance: Maintenance can significantly increase the "productive time" of a system's lifecycle by replacing components or increasing components' lives by eliminating/decreasing the environmental stressors that affect it. One of the potential issues with this approach to risk management, however, is that it usually inserts a human into the loop, thereby making the system more susceptible to negligence or other human errors in the maintenance process.

3. Alarms: Alarms act during the "degradation time" of a system's lifecycle and are useful when the system can tolerate a little bit of change while an intervention is made to impede progress toward an actual failure. Like maintenance, this approach to risk management is also susceptible to human mistakes.

4. Mitigations: These processes lengthen the "degradation time" without having an effect on the consequences of a failure.

5. Controls: Controls act like a safety net. They do not affect the life of a component or the duration of a failure, but minimize the consequences of failures.

6. Containments: Containments do not prevent a failure from occurring but can minimize or eliminate a chain reaction in which one failure causes additional accidents. An example is the containment structure on top of nuclear power plants

7. Design Change: Change is another tool in risk management by which the designer completely changes his or her approach to a problem. Of course, such a step will change the failure scenario and may require a whole new set of risk modelling.

In a risk based design problem, the time interval of the problem under study (productive time, degradation time, failure time, and failure recovery and management time) and the required risk management tools should be identified. Then, system models are developed accordingly.

## 5. Case Study: Satellite Example

For illustrative purposes, an earth observation satellite system is presented as an example. The goal is to consider the early stages of the satellite's design as an example of how the proposed SIMPRA methodology improves the design by making it risk informed. This example highlights the flexibility of the risk scenario planning process in adapting itself to the new system settings after applying design risk management tools.

### 5.1. System Description

An earth observation satellite (EOS) system have several operational modes in a cycle. In each mode, it utilizes its components to perform the tasks. The main operational modes include:

**Receive-command (uplink-data):** In this mode, commands are sent to the satellite for observation and housekeeping purposes to determine the locations and timing of observations and to maintain the satellite in a healthy state. The satellite's communication subsystem (including receiver and antenna), computer, Reaction Control System (RCS), BUS, and software are involved in the functionalities of this mode.

**Collect-data:** The satellite collects the planned data from received commands and store it in the memory. The satellite's RCS, computer, BUS, and software are involved in the functionalities of this mode.

**Process-data:** The collected raw data undergo several processes including compression and quality control for future use. The satellite's BUS, computer, and software are involved in the functionalities of this mode.

**Downlink-data:** The satellite and the ground station can communicate in a limited range. The data collected outside of the limited range are transmitted to the ground as soon as the satellite is in the limited range. The Satellite's transmitter, antenna, computer, BUS, software, and RCS are involved in the functionalities of this mode.

**Standby:** The ground station needs some time to process the transmitted data and provide a new plan for data collection and housekeeping purposes. During this time, the satellite is on Standby mode. The satellite's software and computer and are involved in the functionalities of this mode.

**Safe mode:** It is assumed that satellite enters the Safe mode when there is a problem in the collect-data mode. The satellite's transmitter, antenna, computer, and software are involved in the functionalities of this mode.

**Fail:** The satellite goes through the fail mode when a major component has failed, and the satellite cannot recover. It is assumed that the system enters fail mode from safe modes and collect-data. Thus, the satellite goes to fail mode if software, computer, RCS, or BUS fail.

### 5.2. Design Scenarios

Let us assume that the risk assessment of the satellite design shows that the probability of the system going into the degraded state is higher than the risk acceptability level. At this point the designer studies the scenarios leading to this state and finds out that the actual reason behind this problem is the interaction between software and hardware in the downlink process. The designer determines that one way to solve this problem is to detect the problem early in the downlink process and send the satellite into safe mode for a short period of time to restart the process. Two alternatives for the early detection process are: I) to use an alarm mechanism on the ground, or II) to use a control mechanism on the satellite. The design of the system can then be altered to accommodate these changes:

Option I: An alarm mechanism is designed to check the quality of the data automatically and warn users of possible problems with the data before it is too late to ask the satellite to resend it. The alarm system that is modeled here has three states, DETECT, ACTIVATED and NOT_ACTIVATED. There is also a human decision model added with three states, OBSERVE, SEND_TO_SM and NO_SM. Figure 3 illustrates the changes made in the model to accommodate for applying the alarm system.

Since Alarm and Human are not on the control variable list, their effect is only modeled by detecting their presence. This means that the planner will only generate these scenarios if it detects the presence of these signals in the simulation.

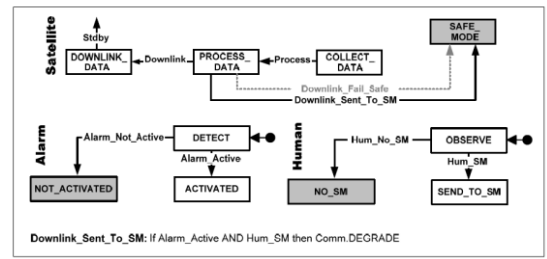

Figure 3. The alarm system added to the satellite system

Option II: The same problem is addressed with a control rather than with an alarm. A quality control unit independently checks the downlink data on the satellite. If it detects a problem with the quality of data, it will presume that the transmitter is degraded so that the system will go to safe mode for repair. Figure 4 illustrates this process.

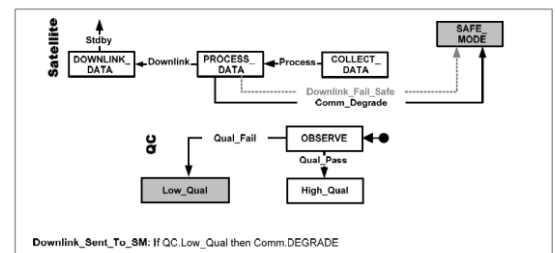

Figure 4. A quality control unit added to the satellite system

New risk scenarios are generated based on these new designs. After running the risk simulation for each alternative, it is observed that although both options can reduce the probability of the degrade state, the use of an alarm on the ground is preferable because of the higher accuracy achieved by having a human checking the quality of transferred data as well.

As the above examples indicate, the risk scenario planning model can accommodate these design risk management changes with a few relatively minor changes. However, these two scenarios have very different effects on the system. One will send the system to Safe Mode with some delay and only after alarming a human, while the other one automatically sends the system to Safe Mode without involving the ground station. One of the major factors for deciding between these risk management tools is how they affect the accuracy of the system as reflected in the results of the simulation itself. The analyst will apply these changes one by one and compare the results of the risk analysis for each of these designs and choose the one that best suits the needs of the system.

## 6. Conclusion

In this paper, a simulation based probabilistic risk assessment (SIMPRA) method is proposed to perform risk-based design for complex systems. One of the main modules of the proposed method is the planner. The planner generates a large number of scenarios automatically that guide the simulation toward different states of the system, and explore unplanned scenarios (particularly useful in identifying new vulnerabilities in the design of the system).

In the risk-based design, the risk of the system for a specific design is evaluated. At this step, end-states, scenarios that lead to the end-states, and probabilities of those end-states are generated.

The system designer will then decide if the risks are acceptable or not. Because of the hierarchical structure of the methodology, in the case that risks are still not acceptable, the designer can easily change the planner and simulation model to manage the risk and enhance the design. This process will be repeated, until the risk is lower than the acceptable levels, specified by designers/experts.

Another application of the proposed risk-based design is to compare multiple available design options. A satellite system is presented as a case study which has two options of alarm system and adding a quality control unit. After performing risk assessment using SIMPRA method, it is concluded that the use of an alarm on the ground is preferable because of the higher accuracy achieved in comparison with installing a quality control unit.

Results of this application illustrate that SIMPRA as a guided risk-based simulation is a very effective tool for risk assessment in the design of large and complex systems.